\documentclass[twocolumn,floatfix,showpacs,superscriptaddress]{revtex4}
\usepackage{graphicx}
\usepackage{amsmath}
\usepackage{amssymb}
\usepackage{bm}

\begin{document}

\title{Majorana bound states without vortices in topological superconductors with electrostatic defects}
\author{M. Wimmer}
\affiliation{Instituut-Lorentz, Universiteit Leiden, P.O. Box 9506, 2300 RA Leiden, The Netherlands}
\author{A. R. Akhmerov}
\affiliation{Instituut-Lorentz, Universiteit Leiden, P.O. Box 9506, 2300 RA Leiden, The Netherlands}
\author{M. V. Medvedyeva}
\affiliation{Instituut-Lorentz, Universiteit Leiden, P.O. Box 9506, 2300 RA Leiden, The Netherlands}
\author{J. Tworzyd{\l}o}
\affiliation{Institute of Theoretical Physics, University of Warsaw, Ho\.{z}a 69, 00--681 Warsaw, Poland}
\author{C. W. J. Beenakker}
\affiliation{Instituut-Lorentz, Universiteit Leiden, P.O. Box 9506, 2300 RA Leiden, The Netherlands}
\date{February 2010}

\begin{abstract}
Vortices in two-dimensional superconductors with broken time-reversal and spin-rotation symmetry can bind states at zero excitation energy. These socalled Majorana bound states transform a thermal insulator into a thermal metal and may be used to encode topologically protected qubits. We identify an alternative mechanism for the formation of Majorana bound states, akin to the way in which Shockley states are formed on metal surfaces: \textit{An electrostatic line defect can have a pair of Majorana bound states at the end points.} The Shockley mechanism explains the appearance of a thermal metal in vortex-free lattice models of chiral \textit{p}-wave superconductors and (unlike the vortex mechanism) is also operative in the topologically trivial phase.
\end{abstract}
\maketitle

Two-dimensional superconductors with spin-polarized-triplet, \textit{p}-wave pairing symmetry have the unusual property that vortices in the order parameter can bind a nondegenerate state with zero excitation energy \cite{Kop91,Vol99,Rea00,Iva01}. Such a midgap state is called a Majorana bound state, because the corresponding quasiparticle excitation is a Majorana fermion --- equal to its own antiparticle. A pair of spatially separated Majorana bound states encodes a qubit, in a way which is protected from any local source of decoherence \cite{Kit01}. Since such a qubit might form the building block of a topological quantum computer \cite{Nay08}, there is an intensive search \cite{Kal09,Tew07,Sat09,Sau09,Lee09,Ali09} for two-dimensional superconductors with the required combination of broken time-reversal and spin-rotation symmetries (symmetry class \textit{D} \cite{Alt97}).

The generic Bogoliubov-De Gennes Hamiltonian $H$ of a chiral \textit{p}-wave superconductor is only constrained by particle-hole symmetry, $\sigma_{x}H^{\ast}\sigma_{x}=-H$. At low excitation energies $E$ (to second order in momentum $\bm{p}=-i\hbar\partial/\partial\bm{r}$) it has the form
\begin{equation}
H=\Delta\bigl(p_{x}\sigma_{x}+p_{y}\sigma_{y}\bigr)+\bigl(U(\bm{r})+p^{2}/2m\bigr)\sigma_{z},\label{Hdef}
\end{equation}
for a uniform (vortex-free) pair potential $\Delta$. The electrostatic potential $U$ (measured relative to the Fermi energy) opens up a band gap in the excitation spectrum. At $U=0$ the superconductor has a topological phase transition (known as the thermal quantum Hall effect) between two localized phases, one with and one without chiral edge states \cite{Vol88,Sen00,Vis01,Vol03}.

\begin{figure}[htb]
\centerline{\includegraphics[width=0.8\linewidth]{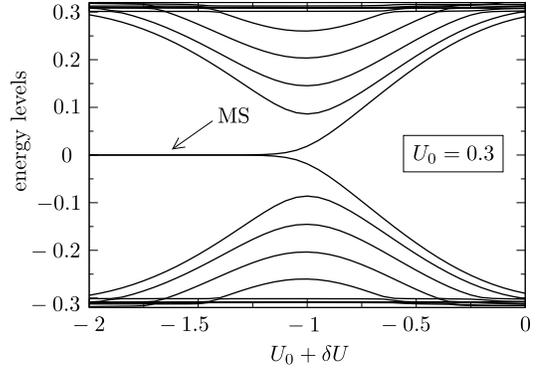}}
\caption{\label{fig_spectrum}
Emergence of a pair of zero-energy MS states as the defect potential $U_{0}+\delta U$ is made more and more negative, at fixed positive background potential $U_{0}=0.3$. (All energies are in units of $\gamma\equiv\hbar\Delta/a$.) The energy levels are the eigenvalues of the Hamiltonian \eqref{Hdef} on a square lattice (dimension $100\,a\times 100\,a$, $\beta\equiv\hbar^{2}/2ma^{2}=0.4\,\gamma$, periodic boundary conditions). The line defect has length $50\,a$. The dense spectrum at top and bottom consists of bulk states.
}
\end{figure}

Our key observation is that the Hamiltonian \eqref{Hdef} on a lattice has Majorana bound states at the two end points of a linear electrostatic defect (consisting of a perturbation of $U$ on a string of lattice sites). The mechanism for the production of these bound states goes back to Shockley \cite{Sho39}: The band gap closes and then reopens upon formation of the defect, and as it reopens a pair of states splits off from the band edges to form localized states at the end points of the defect (see Fig.\ \ref{fig_spectrum}). Such Shockley states appear in systems as varied as metals and narrow-band semiconductors \cite{Dav96}, carbon nanotubes \cite{Sav09}, and photonic crystals \cite{Mal09}. In these systems they are unprotected and can be pushed out of the band gap by local perturbations. In a superconductor, in contrast, particle-hole symmetry requires the spectrum to be $\pm E$ symmetric, so an isolated bound state is constrained to lie at $E=0$ and cannot be removed by any local perturbation. 

We propose the name Majorana-Shockley (MS) bound state for this special type of topologically protected Shockley states. Similar states have been studied in the context of lattice gauge theory by Creutz and Horv\'{a}th \cite{Cre94,Cre01}, for an altogether different purpose (as a way to restore chiral symmetry in the Wilson fermion model of QCD \cite{Wil74}).

Consider a square lattice (lattice constant $a$), at uniform potential $U_{0}$. The Hamiltonian \eqref{Hdef} on the lattice has dispersion relation
\begin{align}
E^{2}={}&[U_{0}+2\beta(2-\cos ak_{x}-\cos ak_{y})]^{2}\nonumber\\
&+\gamma^{2}\sin^{2}ak_{x}+\gamma^{2}\sin^{2}ak_{y}.\label{Ekxky}
\end{align}
(We have defined the energy scales $\beta=\hbar^{2}/2ma^{2}$, $\gamma=\hbar\Delta/a$.) The spectrum becomes gapless for $U_{0}=0$, $-4\beta$, and $-8\beta$, signaling a topological phase transition \cite{Qi06}. The number of edge states is zero for $U_{0}>0$ and $U_{0}<-8\beta$, while it is unity otherwise (with a reversal of the direction of propagation at $U_{0}=-4\beta$). The topologically nontrivial regime is therefore reached for $U_{0}$ negative but larger than $-8\beta$.

\begin{figure}[tb]
\centerline{\includegraphics[width=0.9\linewidth]{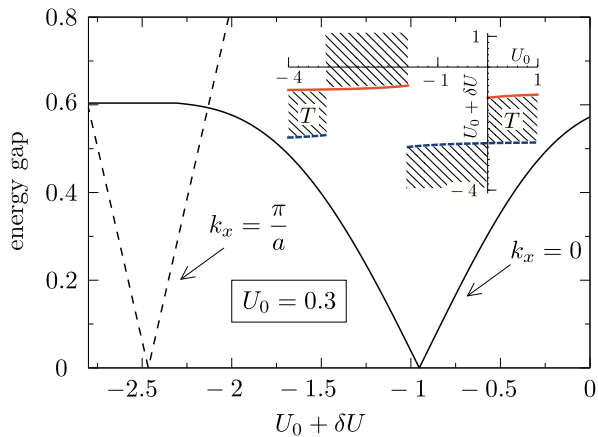}}
\caption{\label{fig_gap}
Main plot: Closing and reopening of the excitation gap at $U_{0}=0.3$, $\beta=0.4$ (in units of $\gamma$), for states with $k_{x}=0$ (black solid curve) and $k_{x}=\pi/a$ (black dashed curve). The MS states exist for defect potentials in between two gap-closings, indicated as a function of $U_{0}$ by the shaded regions in the inset. (The red solid and blue dashed curves show, respectively $U_{0}+\delta U_{0}$ and $U_{0}+\delta U_{\pi}$. The label $T$ indicates the topologically trivial phase.)
}
\end{figure}

We now introduce the electrostatic line defect by changing the potential to $U_{0}+\delta U$ on the $N$ lattice points at $\bm{r}=(na,0)$, $n=1,2,\ldots N$. In Figs.\ \ref{fig_spectrum} and \ref{fig_gap} we show the closing and reopening of the band gap as the defect is introduced, accompanied by the emergence of a pair of states at zero energy. The eigenstates for which the gap closes and reopens have wave vector $k_{x}$ parallel to the line defect equal to either $0$ or $\pm\pi/a$ (in the limit $N\rightarrow\infty$ when $k_{x}$ is a good quantum number). 

We have calculated that the gap closing at $k_{x}=0$ happens at a critical potential $\delta U=\delta U_{0}$ given by \cite{appendix}
\begin{equation}
\delta U_{0}=\begin{cases}
-\sqrt{U_{0} (U_{0}+4 \beta)+\gamma^{2}}&\text{for $U_{0}>0$},\\
\sqrt{U_{0} (U_{0}+4 \beta)+\gamma^{2}}&\text{for $U_{0}<-4\beta$},\\
\text{no finite value otherwise}.& 
\end{cases}\label{Uk0result}
\end{equation}
The critical potential $\delta U_{\pi}$ for closing of the gap at $k_{x}=\pm\pi/a$ is obtained from Eq.\ \eqref{Uk0result} by the replacement of $U_{0}$ with $U_{0}+4\beta$. The MS states appear for defect potentials $U_{0}+\delta U$ in between two subsequent gap closings, as indicated in the inset of Fig.\ \ref{fig_gap}.

We conclude that MS states exist for any value of $U_{0}$. In contrast, Majorana bound states in vortices exist only in the topologically nontrivial regime \cite{Rea00,Gur07}. The index theorem \cite{Roy10} for the production of zero-energy modes by the vortex mechanism, which requires the topologically nontrivial phase, is therefore not applicable to the Shockley mechanism.

\begin{figure}[tb]
\centerline{\includegraphics[width=0.95\linewidth]{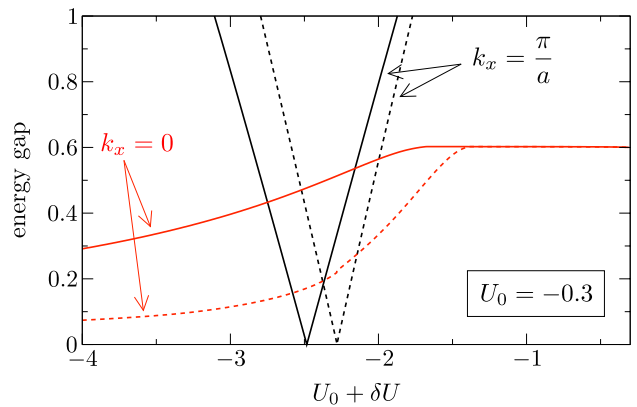}}
\caption{\label{selfconsistency}
Closing and reopening of the excitation gap at $U_{0}=-0.3$, $\beta=0.4$ (in units of $\gamma$), for states with $k_{x}=0$ (red curves) and $k_{x}=\pi/a$ (black curves).
The results were obtained from numerical calculations using a constant isotropic pair potential $\Delta$ (solid lines) as in Fig.~\ref{fig_gap}
as well as a spatially dependent, anisotropic pair potential $(\Delta_x(\bm{r}), \Delta_y(\bm{r}))$ determined self-consistently from the gap equation (dashed lines) \cite{appendix}.}
\end{figure}

Our reasoning so far has relied on the assumption of a constant pair potential $\Delta$, unperturbed by the defect. In order to demonstrate the robustness of the Majorana-Shockley mechanism, we have performed numerical calculations that determine the pair potential self-consistently by means of the gap equation \cite{appendix, Furusaki01}. In Fig.~\ref{selfconsistency} we show a comparison of the closing and reopening
of the band gap as obtained from calculations with and without self-consistency, in the relevant weak pairing regime ($U_0<0$).  The
self-consistency does not change the qualitative behavior. In particular, the gap only closes at $k_x=\pi/a$ for the parameters
chosen (c.f.~inset in Fig.~\ref{fig_gap}) and the self-consistent determination of $\Delta$ only shifts the critical potential $\delta U$ slightly.

\begin{figure}[tb]
\centerline{\includegraphics[width=0.7\linewidth]{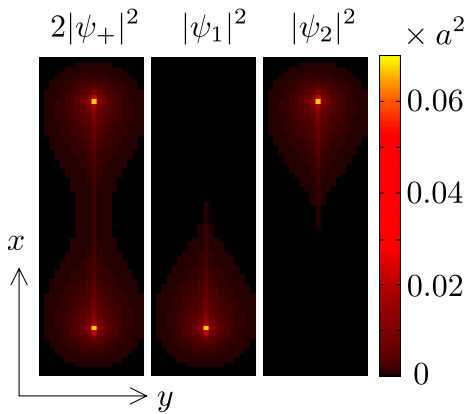}}
\caption{\label{fig_wavefunction}
Probability density of the paired ($\psi_{+}$) and unpaired ($\psi_{1},\psi_{2}$) Majorana bound states at the end points of a line defect of length $50\,a$, calculated for $U_{0}=0.1\,\gamma$, $U_{0}+\delta U=-1.3\,\gamma$, $\beta=0.4\,\gamma$.
}
\end{figure}

In Fig.\ \ref{fig_wavefunction} we demonstrate that the MS states are localized at the end points of the line defect. The exponentially small, but nonzero overlap of the pair of states displaces their energy from $0$ to $\pm E$ (with corresponding eigenstates $\psi_{-}=\sigma_{x}\psi_{+}^{\ast}$ related by particle-hole symmetry). The unpaired Majorana bound states $\psi_{1}$ and $\psi_{2}$ are given by the linear combinations
\begin{subequations}
\label{psi12}
\begin{align}
&\psi_{1}=\tfrac{1}{2}(1-i)\psi_{+}+\tfrac{1}{2}(1+i)\psi_{-},\label{psi1}\\
&\psi_{2}=\tfrac{1}{2}(1+i)\psi_{+}+\tfrac{1}{2}(1-i)\psi_{-},\label{psi2}
\end{align}
\end{subequations}
shown also in Fig.\ \ref{fig_wavefunction}. These states are particle-hole symmetric, $\psi_{1,2}=\sigma_{x}\psi_{1,2}^{\ast}$, so the quasiparticle in such a state is indeed equal to its own antiparticle (hence, it is a Majorana fermion).

\begin{figure}[tb]
\centerline{\includegraphics[width=0.9\linewidth]{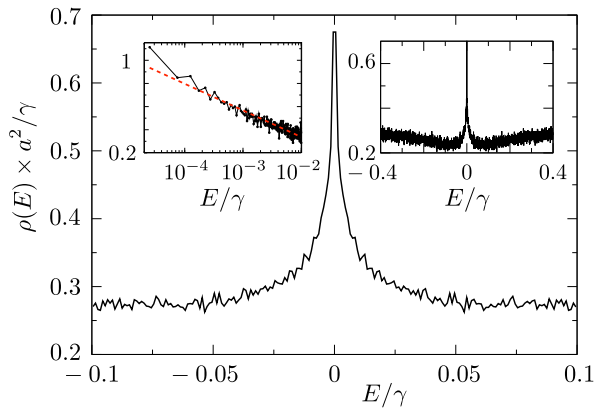}}
\caption{\label{fig_peak}
Average density of states for a potential that fluctuates randomly from site to site ($\bar{U}=0.01\,\gamma$, $\Delta U=2\,\gamma$, $\beta=0.2\,\gamma$). The lattice has size $400\,a\times 400\,a$. The right inset shows the same data as in the main plot, over a larger energy range. The left inset has a logarithmic energy scale, to show the dependence $\rho\propto\ln|E|$ expected for a thermal metal (red dashed line).
}
\end{figure}

If the line defect has a width $W$ which extends over several lattice sites, multiple gap closings and reopenings appear at $k_{x}=0$ upon increasing the defect potential $U_{0}+\delta U\equiv-(\hbar k_{F})^{2}/2m$ to more and more negative values at fixed positive background potential $U_{0}$. In the continuum limit $W/a\rightarrow\infty$, the gap closes when \cite{appendix} $qW=n\pi+\nu$, $n=0,1,2,\ldots$, with $q=[k_{F}^{2}-(m\Delta)^{2}]^{1/2}$ the real part of the transverse wave vector and $\nu\in(0,\pi)$ a phase shift that depends weakly on the potential. (Similar oscillatory coupling energies of zero-modes have been found in Refs.\ \cite{Che09,Liu10}.) The MS states at the two ends of the line defect alternatingly appear and disappear at each subsequent gap closing.

So far we constructed MS states for a linear electrostatic defect. More generally, we expect a randomly varying electrostatic potential to create a random arrangement of MS states. To test this, we pick $U(\bf{r})$ at each lattice point uniformly from the interval $(\bar{U}-\Delta U,\bar{U}+\Delta U)$ and calculate the average density of states $\rho(E)$. The result in Fig.\ \ref{fig_peak} shows the expected peak at $E=0$. This peak is characteristic of a thermal metal, studied previously in models where the Majorana bound states are due to vortices \cite{Boc00,Cha02,Mil07}. The theory of a thermal metal \cite{Sen00} predicts a logarithmic profile, $\rho(E)\propto\ln |E|$, for the peak in the density of states, which is consistent with our data.

Without Majorana bound states, the chiral \textit{p}-wave superconductor would be in the thermal insulator phase, with an exponentially small thermal conductivity at any nonzero $\bar{U}$ \cite{Rea00,Boc00,Rea01,Bar10}.  Our findings imply that electrostatic disorder can convert the thermal insulator into a thermal metal, thereby destroying the thermal quantum Hall effect. Numerical results for this insulator-metal transition will be reported elsewhere \cite{Med10}.

These results are all for a specific model of a chiral \textit{p}-wave superconductor. We will now argue that our findings are generic for symmetry class \textit{D} (along the lines of a similar analysis of solitons in a polymer chain \cite{Jac83}). Let $p$ be the momentum along the line defect and $\alpha$ a parameter that controls the strength of the defect. Assume that the gap closes at $\alpha=\alpha_{0}$ and at $p=0$. (Because of particle-hole symmetry the gap can only close at $p=0$ or $p=\pm\hbar\pi/a$ and these two cases are equivalent.) For $\alpha$ near $\alpha_{0}$ and $p$ near $0$ the Hamiltonian in the basis of left-movers and right-movers has the generic form
\begin{equation}
H(\alpha)=\begin{pmatrix}
(v_{0}+v_{1})p&-i(\alpha-\alpha_{0})\\
i(\alpha-\alpha_{0})&-(v_{0}-v_{1})p
\end{pmatrix},\label{Halpha}
\end{equation}
with velocities $0<v_{1}<v_{0}$. No other terms to first order in $p=-i\hbar\partial/\partial x$ and $\alpha-\alpha_{0}$ are allowed by particle-hole symmetry, $H(\alpha)=-H^{\ast}(\alpha)$.

The line defect is initially formed by letting $\alpha$ depend on $x$ on a scale much larger than the lattice constant. We set one end of the defect at $x=0$ and increase $\alpha$ from $\alpha(-\infty)<\alpha_{0}$ to $\alpha(+\infty)>\alpha_{0}$. Integration of $H[\alpha(x)]\psi(x)=0$ then gives the wave function of a zero-energy state bound to this end point,
\begin{equation}
\psi(x)=\begin{pmatrix}
\sqrt{v_{0}/v_{1}-1}\\
\sqrt{v_{0}/v_{1}+1}
\end{pmatrix}\exp\left(-\int_{0}^{x} \frac{\alpha(x')-\alpha_{0}}{\sqrt{v_{0}^{2}-v_{1}^{2}}} \,dx'\right).\label{psiMS}
\end{equation}
This is one of the two MS states, the second being at the other end of the line defect. We may now relax the assumption of a slowly varying $\alpha(x)$, since a pair of uncoupled zero-energy states cannot disappear without violating particle-hole symmetry.

We conclude with an outlook. We have identified a purely electrostatic mechanism for the creation of Majorana bound states in chiral \textit{p}-wave superconductors. The zero-energy (mid-gap) states appear in much the same way as Shockley states in non-superconducting materials, but now protected from any local perturbation by particle-hole symmetry. An experimentally relevant consequence of our findings is that the thermal quantum Hall effect is destroyed by electrostatic disorder (in marked contrast to the electrical quantum Hall effect). A recent proposal to realize Wilson fermions in optical lattices \cite{Bermudez10} also opens the possibility to observe Majorana-Shockley states using cold atoms.

Our analysis is based on a generic model of a two-dimensional class-\textit{D} superconductor (broken time-reversal and spin-rotation symmetry). An interesting direction for future research is to explore whether Majorana-Shockley bound states exist as well in the other symmetry classes \cite{Alt97}. Since an electrostatic defect preserves time-reversal symmetry, we expect the Majorana-Shockley mechanism to be effective also in class \textit{DIII} (when only spin-rotation symmetry is broken). That class includes proximity-induced \textit{s}-wave superconductivity at the surface of a topological insulator \cite{Fu08} and other experimentally relevant topological superconductors \cite{Qi09,Sch09,Fu09}.

It would also be interesting to investigate the braiding of two electrostatic defect lines, in order to see whether one obtains the same non-Abelian statistics as for the braiding of vortices \cite{Iva01}.
 
We have benefited from discussions with B. B\'{e}ri, L. Fu, and C.-Y. Hou. This research was supported by the Deutscher Akademischer Austausch Dienst DAAD, by the Dutch Science Foundation NWO/FOM, by an ERC Advanced Investigator Grant, and by the EU network NanoCTM.

\appendix
\section{Line defect in lattice fermion models}
\label{linedefect}

We calculate the closing and reopening of the excitation gap upon introduction of a line defect in a lattice fermion model with particle-hole symmetry. First we treat the Wilson fermion model \cite{Wil74} considered in the main text, and introduced in the context of topological insulators in Refs.\ \cite{Ber06,Fu07}. Then, in order to demonstrate the generic nature of the results, we consider an alternative lattice model, the staggered fermion (or Kogut-Susskind) model \cite{Kog75,Sta82,Ben83}, introduced in the context of graphene in Refs.\ \cite{Two08,Med10}.

\subsection{Wilson fermions}
\label{Wilson}

The Wilson fermion model has Hamiltonian
\begin{equation}
H=\sum_{n}c^{\dagger}_{n}{\cal E}_{n}c^{\vphantom{\dagger}}_{n}-\!\!\!\!\sum_{n,m\;{\rm (nearest\; neighb.)}}c_{n}^{\dagger}{\cal T}_{nm}c^{\vphantom{\dagger}}_{m}.\label{Htb}
\end{equation}
Each site $n$ on a two-dimensional square lattice (lattice constant $a$) has electron and hole states $|e\rangle$ and $|h\rangle$. Fermion annihilation operators for these two states are collected in a vector $c_{n}=(c_{n,e},c_{n,h})$. States on the same site are coupled by the $2\times 2$ potential matrix ${\cal E}_{n}$ and states on adjacent sites by the $2\times 2$ hopping matrix ${\cal T}_{nm}$, defined by \cite{Ber06,Fu07}
\begin{equation}
{\cal E}_{n}=\begin{pmatrix}
U_{n}&0\\
0&-U_{n}
\end{pmatrix},\;\;
{\cal T}_{nm}=\begin{pmatrix}
\beta &\gamma e^{i\theta_{nm}}\\
\gamma e^{-i\theta_{mn}}&-\beta
\end{pmatrix}.\label{ETdef}
\end{equation}
Here $U_n$ is the electrostatic potential on site $n$ and $\theta_{nm}\in[0,\pi]$ is the angle between the vector $\bm{r}_{n}-\bm{r}_{m}$ and the positive $y$-axis (so $\theta_{mn}=\pi-\theta_{nm}$). In the continuum limit $a\rightarrow 0$, the tight-binding Hamiltonian \eqref{Htb} is equivalent to the chiral \textit{p}-wave Hamiltonian \eqref{Hdef}, with $\beta=\hbar^{2}/2ma^{2}$ and $\gamma=\hbar\Delta/a$.

It is convenient to transform from position to momentum representation. For that purpose we take periodic boundary conditions in the $y$-direction, so that the transverse wavevector (in units of $1/a$) has the discrete values $k_{l}=2\pi l/N$, $l=-(N-1)/2,\ldots,-1,0,1,\ldots,(N-1)/2$ (for an odd number $N$ of sites in the $y$-direction). The Fourier transformation from position to momentum representation is carried out by the unitary matrix with elements $[{\cal F}]_{nl}=N^{-1/2}e^{ink_l}$. We take an infinitely long system in the $x$-direction, so the longitudinal wavevector $k$ varies continuously in the interval $(-\pi,\pi]$. 

For a uniform potential, $U_{n}\equiv U_{0}$ for all $n$, the Fourier transformed Hamiltonian $H_{0}(k)$ has matrix elements
\begin{align}
[H_{0}(k)]_{ll'}={}&\delta_{ll'}{\cal E}_{l}(k),\label{H0E}\\
{\cal E}_{l}(k)={}&U_{0}\sigma_{z}+2\beta\sigma_{z}(2-\cos k-\cos k_l)\nonumber\\
&+\gamma(\sigma_x\sin k+\sigma_y\sin k_l).\label{Edef}
\end{align}
The corresponding dispersion relation is
\begin{align}
E(k,k_{l})^{2}={}&[U_{0}+2\beta(2-\cos k-\cos k_{l})]^{2}\nonumber\\
&+\gamma^{2}(\sin^{2}k+\sin^{2}k_{l}),\label{Ekkl}
\end{align}
cf.\ Eq.\ \eqref{Ekxky}. 

A line defect at row $n_{0}$ (parallel to the $x$-axis) adds to $H_{0}$ the perturbation
\begin{equation}
[\delta H]_{ll'}=N^{-1} e^{in_0(k_{l'}-k_{l})}\delta U\sigma_{z}.\label{deltaH}
\end{equation}
The determinantal equation ${\rm Det}\,(H_{0}+\delta H-E)=0$ for eigenenergy $E$ reads
\begin{equation}
{\rm Det}\,(1+{\cal F}_{0}^{\dagger}\delta U\sigma_{z}{\cal F}_{0}(H_{0}-E)^{-1})=0,\label{detE1}
\end{equation}
in terms of an $1\times N$ matrix ${\cal F}_{0}$ with elements $[{\cal F}_{0}]_{1l}=N^{-1/2}e^{in_{0}k_{l}}$. Sylvester's theorem, ${\rm Det}(1+AB)={\rm Det}(1+BA)$, allows us to rewrite the determinant in the form
\begin{equation}
{\rm Det}\,(1+\delta U\sigma_{z}{\cal F}_{0}(H_{0}-E)^{-1}{\cal F}_{0}^{\dagger})=0,\label{detE2}
\end{equation}
which reduces to
\begin{align}
0&=
{\rm Det}\,\left(1+\delta U\sigma_{z}\frac{1}{N}\sum_{l}\frac{1}{{\cal E}_{l}(k)-E}\right)\nonumber\\
&={\rm Det}\,\left(1+\delta U\sigma_{z}\frac{1}{N}\sum_{l}\frac{{\cal E}_{l}(k)+E}{E(k,k_{l})^{2}-E^{2}}\right).\label{detE3}
\end{align}

A zero-mode is a pair of states (one left-mover and one right-mover) at energy $E=0$. This can only occur at $k=0$ or $k=\pi$ (because for any eigenenergy $E$ at $k$ there must also be an eigenenergy $-E$ at $-k$). From Eqs.\ \eqref{Edef} and \eqref{detE3} we obtain the condition for such a zero-mode,
\begin{equation}
\frac{1}{N}\sum_{l}\frac{U_{0}+2\beta(1+\delta-\cos k_{l})}{[U_{0}+2\beta(1+\delta-\cos k_{l})]^{2}+\gamma^{2}\sin^{2}k_{l}}=-\frac{1}{\delta U},\label{detE4}
\end{equation}
where $\delta=0$ if $k=0$ and $\delta=2$ if $k=\pi$. In the limit $N\rightarrow\infty$ we may replace the sum by an integral, $N^{-1}\sum_{l}\rightarrow(2\pi)^{-1}\int_{-\pi}^{\pi}dk_{l}$, which can be evaluated by contour integration. The resulting critical value of $\delta U$ is given in the main text [Eq.\ \eqref{Uk0result} and following].

\subsection{Staggered fermions}
\label{staggered}

The staggered fermion model is a discretization of the Hamiltonian \eqref{Hdef} without the $p^{2}$ term. It is formulated in Refs.\ \cite{Sta82,Ben83,Two08} in terms of the transfer matrix ${\cal M}_{m}$, which relates the transverse wave functions $\Psi_{m+1}={\cal M}_{m}\Psi_{m}$ at columns $m$ and $m+1$ (parallel to the $y$-axis). For a line defect along the $x$-axis, the transfer matrix is $m$-independent, so we can omit the column number $m$. 

The transfer matrix (at energy $E$) has the form
\begin{align}
&{\cal M}=\frac{1-iX}{1+iX},\label{Tdef}\\
&X=(\gamma{\cal J})^{-1}(\gamma\sigma_{z}{\cal K}+\tfrac{1}{2}E\sigma_{x}{\cal J}-\tfrac{1}{2}i\sigma_{y}{\cal U}).\label{Xdef}
\end{align}
In reference to Eq.\ \eqref{Hdef}, the parameter $\gamma=\hbar\Delta/a$ for lattice constant $a$. The $N\times N$ matrices ${\cal J}$ and ${\cal K}$ have nonzero elements
\begin{align}
&{\cal J}_{n,n}=1,\;\;{\cal J}_{n,n+1}={\cal J}_{n,n-1}=\tfrac{1}{2},\label{calIdef}\\
&{\cal K}_{n,n+1}=\tfrac{1}{2},\;\;{\cal K}_{n,n-1}=-\tfrac{1}{2},\label{calKdef}
\end{align}
while the potential matrix ${\cal U}$ (for a line defect at row $n_{0}$) is given by
\begin{align}
{\cal U}_{nn'} ={}& U_0 {\cal J}_{nn'}+\tfrac{1}{2}\delta U (\delta_{n,n'}\delta_{n,n_0}+\delta_{n,n'}\delta_{n,n_0+1}\nonumber\\
&+\delta_{n+1,n'}\delta_{n,n_0}+\delta_{n,n'+1}\delta_{n',n_0}).\label{calUdef}
\end{align}

In momentum representation, the matrix $X$ has elements
\begin{equation}
X_{ll'}={\cal A}_{l}\delta_{ll'}-i(\delta U/2\gamma)\sigma_{y}\frac{v^{*}_{l}v^{\vphantom{*}}_{l'}}{4\cos^{2}(k_l/2)},\label{Xmomentum}
\end{equation}
where we have defined
\begin{align}
&{\cal A}_{l}=i\sigma_{z}\tan(k_{l}/2)+(E/2\gamma)\sigma_{x}-i(U_{0}/2\gamma)\sigma_{y},\label{calAdef}\\
&v_{l}=N^{-1/2}e^{in_{0}k_{l}}(1+e^{ik_{l}}).\label{vldef}
\end{align}
The dispersion relation of the staggered fermions is $\tan^{2}(k/2)=A(k,k_{l})^{2}$, with
\begin{equation}
A(k,k_{l})^{2}=(E/2\gamma)^{2}-\tan^{2}(k_{l}/2)-(U_{0}/2\gamma)^{2}.\label{Akkldef}
\end{equation}

An eigenstate at energy $E$ and longitudinal wavevector $k$ is an eigenstate of $X$ with eigenvalue $-\tan(k/2)$. The determinantal equation ${\rm Det}[X+\tan(k/2)]=0$ can again be simplified using Sylvester's theorem. The result, analogous to Eq.\ \eqref{detE3}, is
\begin{align}
0&={\rm Det}\,\left(1-\frac{\delta U}{2\gamma}i\sigma_{y}\frac{1}{N}\sum_{l}\frac{1}{{\cal A}_{l}+\tan(k/2)}\right)\nonumber\\
&={\rm Det}\,\left(1-\frac{\delta U}{2\gamma}i\sigma_{y}\frac{1}{N}\sum_{l}\frac{{\cal A}_{l}-\tan(k/2)}{A(k,k_{l})^{2}-\tan^{2}(k/2)}\right).\label{detE5}
\end{align}

\begin{figure}[tb]
\centerline{\includegraphics[width=0.8\linewidth]{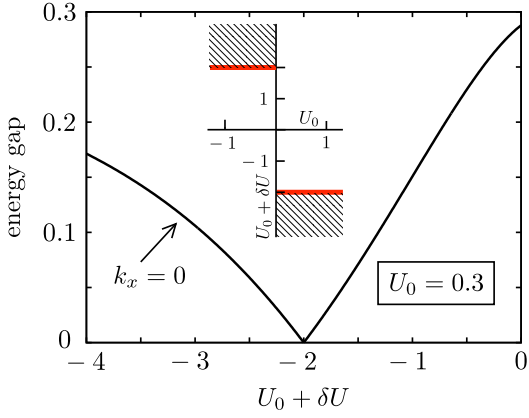}}
\caption{\label{fig_gap_staggered}
Main plot: Closing and reopening of the excitation gap in the staggered fermion model. The MS states exist for defect potentials in the shaded regions in the inset. (All energies are in units of $\gamma$.)
}
\end{figure}

Because of the pole in the dispersion relation at $k=\pi$, the zero-mode now exists only at $k=0$. The condition for this zero-mode, analogous to Eq.\ \eqref{detE4}, is
\begin{equation}
\frac{1}{N}\sum_{l}\frac{U_{0}/2\gamma}{(U_{0}/2\gamma)^{2}+\tan^{2}(k_{l}/2)}=-\frac{2\gamma}{\delta U},\label{detE6}
\end{equation}
For $N\rightarrow\infty$ we may again transform the sum into an integral, and thus obtain the critical potential
\begin{equation}
\delta U=\begin{cases}
-U_{0}-2\gamma&{\rm if}\;\;U_{0}>0,\\
-U_{0}+2\gamma&{\rm if}\;\;U_{0}<0.
\end{cases}\label{U0staggered}
\end{equation}

Upon varying the potential $U_{0}+\delta U$ of the line defect, at fixed bulk potential $U_{0}$, the closing and reopening of the gap thus happens at $U_{0}+\delta U=-2\gamma\,{\rm sign}\,(U_{0})$ (see Fig.\ \ref{fig_gap_staggered}). The inset shows the region in parameter space where the Majorana-Shockley states exist in the staggered fermion model. This phase diagram is much simpler than the corresponding phase diagram for Wilson fermions (Fig.\ \ref{fig_gap}, inset), because of the absence of the extra parameter $\beta$ (which quantifies the strength of the $p^{2}$ term in the Wilson fermion model).

\section{Self-consistent determination of the pair potential}
\label{app_selfconsistency}

In order to determine the pair potential self-consistently in a spatially non-homogeneous situation, it is necessary to allow for a position-dependent, anisotropic pair potential $\bm{\Delta}(\bm{r})=(\Delta_x(\bm{r}), \Delta_y(\bm{r}))$. The Hamiltonian then reads \cite{Furusaki01}
\begin{align}
H=&\tfrac{1}{2}\left\{\Delta_x(\bm{r}), p_{x}\right\}\sigma_{x}+\tfrac{1}{2}\left\{\Delta_y(\bm{r}), p_{y}\right\}\sigma_{y}\nonumber\\
&+\bigl(U(\bm{r})+p^{2}/2m\bigr)\sigma_{z},\label{app_Hdef}
\end{align} 
where $\{\cdot,\cdot\}$ denotes the anticommutator.
In the discretization of this Hamiltonian on a square lattice, the spatial dependence of $\bm{\Delta}(\bm{r})$ is taken into account in the hopping between neighbors as an average value of $\bm{\Delta}(\bm{r})$ on the two lattice points.

When the pair potential is homogeneous, the lattice Hamiltonian has the 
spectrum
\begin{align}
E^{2}={}&[U_{0}+2\beta(2-\cos ak_{x}-\cos ak_{y})]^{2}\nonumber\\
&+\gamma_x^{2}\sin^{2}ak_{x}+\gamma_y^{2}\sin^{2}ak_{y}\label{app_Ekxky}
\end{align}
with $\gamma_x=\hbar \Delta_x/a$, $\gamma_y=\hbar \Delta_y/a$
and $\beta=\hbar^2/2 m a^2$. 

The Hamiltonian must be solved self-consistently together with the
equation for the pair potential. These read \cite{Furusaki01} (with derivatives discretized on the lattice)
\begin{align}
\gamma_x(\bm{r})&=-i g \sum_{E_n>0} (u_n(x+a,y)-u_n(x-a,y))\, v_n^*(x,y)\nonumber\\
&\qquad -u_n(x,y)\, (v_n^*(x+a,y)-v_n^*(x-a,y)),\nonumber\\
\gamma_y(\bm{r})&=g \sum_{E_n>0} (u_n(x,y+a)-u_n(x,y+a))\, v_n^*(x,y)\nonumber\\
&\qquad -u_n(x,y)\, (v_n^*(x,y+a)-v_n^*(x,y+a)).
\end{align}
Here $u_n$ and $v_n$ are the electron and hole component of the wave function,
respectively, and assumed to be from the tight-binding model, i.e.
they are dimensionless and represent the probability amplitude per 
lattice point $(x,y)$. 

The coupling constant $g$ must be chosen such that it gives the correct pair potential $\gamma$ in the bulk. It can be calculated as
\begin{align}
\frac{\gamma}{g}=&\frac{1}{\pi^2} \int_{-\pi}^{\pi} d(ak_x) 
\int_{-\pi}^{\pi} d(ak_y)\, \sin(ak_x)\, u(\bm{k}) v^*(\bm{k})
\nonumber\\=&\frac{-i}{\pi^2} \int_{-\pi}^{\pi} d(ak_x) 
\int_{-\pi}^{\pi} d(ak_y)\, \sin(ak_y)\, u(\bm{k}) v^*(\bm{k}),
\end{align}
where $u(\bm{k})$ and $v(\bm{k})$ are the electron and hole 
coefficients of the plane wave solutions of the bulk lattice
Hamiltonian with $E>0$. 

In the particular case of a system that is translationally invariant in $x$-direction, as is the case for an infinitely extended line defect, the gap equations can be written as:
\begin{align}
\gamma_x(\bm{r})&=\frac{4 g}{N_x} \sum_{E_n>0, k_x} u_n(k_x,y) v_n^*(k_x,y)\,
\sin(a k_x)\nonumber\\
\gamma_y(\bm{r})&=\frac{g}{N_x} \sum_{E_n>0, k_x}\nonumber \\ 
&\qquad \biggl( (u_n(k_x,y+a)-u_n(k_x,y+a))\, v_n^*(k_x,y)\nonumber\\
&\qquad -u_n(k_x,y)\, (v_n^*(k_x,y+a)-v_n^*(k_x,y+a)) \biggr),\label{app_gap_kx}
\end{align}
summing over $N_x$ longitudinal momenta $k_x$, and solving the 
tight-binding problem for each $k_x$ individually.

The self-consistent solution of the tight-binding Hamiltonian and the gap equation \eqref{app_gap_kx} is obtained in an iterative procedure. In the iteration, we neglect the influence of the vector potential arising from local currents \cite{Furusaki01} as those effects are expected to be minor for the examples considered in this work. Furthermore, we also avoid adjusting the chemical potential $U_0$  to obtain a fixed number of electrons in
the system and instead use a large unit cell so that the bulk value of $\bm{\Delta}$ is recovered away from the defect.

\section{Line defect in the continuum limit}
\label{continuum}

We calculate the closing and reopening of the excitation gap upon introduction of a line defect in the Hamiltonian \eqref{Hdef}, which is the continuum limit ($a\rightarrow 0$) of the Wilson fermion lattice model of App.\ \ref{Wilson}. The mode matching calculation presented here is the one-dimensional version of the two-dimensional calculation in Refs.\ \cite{Liu10,Lin09,Lu10}.

The line defect, of width $W$, is formed by the electrostatic potential profile
\begin{equation}
U(\bm{r})=\begin{cases}
U_{0}&{\rm if}\;\;|y|>W/2,\\
U_{0}+\delta U&{\rm if}\;\;|y|<W/2.
\end{cases}\label{UWdef}
\end{equation}
A zero-mode $\psi=(u,v)$ is a (doubly degenerate) eigenstate of the Hamiltonian \eqref{Hdef} at $E=0$, $p_{x}=0$. The zero-mode should thus satisfy
\begin{subequations}
\label{uvequations}
\begin{align}
&(U+p_{y}^{2}/2m)u=i\Delta p_{y}v,\label{uveqa}\\
&(U+p_{y}^{2}/2m)v=i\Delta p_{y}u.\label{uveqb}
\end{align}
\end{subequations}
For uniform $U$ the solution is a plane wave,
\begin{equation}
\psi_{ss'}=e^{ik_{ss'}y}\begin{pmatrix}
1\\ s
\end{pmatrix},\;\;s,s'=\pm 1,\label{psisol}
\end{equation}
with transverse wave vector
\begin{equation}
k_{ss'}=(m/\hbar)\bigl(is\Delta+s'\sqrt{-\Delta^{2}-2U/m}\bigr).\label{kssdef}
\end{equation}

\begin{figure}[b]
\centerline{\includegraphics[width=0.7\linewidth]{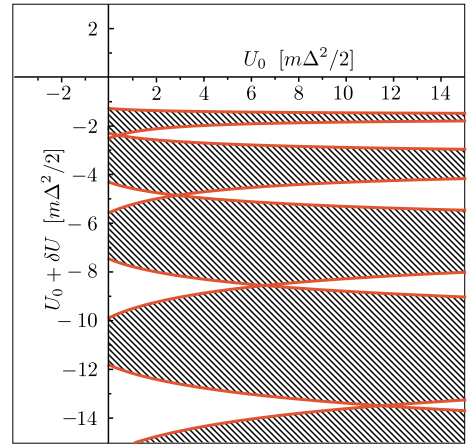}}
\caption{\label{fig_cont}
The red solid curves are the solution of Eq.\ \eqref{qWrelation} for $W=4\hbar/m\Delta$. The MS states exist in the shaded regions.
}
\end{figure}

In the region $|y|<W/2$ the zero-mode $\psi$ is a superposition of the four states $\psi_{++},\psi_{+-},\psi_{-+},\psi_{--}$. For $y>W/2$ two decaying states with ${\rm Im}\,k_{ss'}>0$ appear in the superposition, while for $y<-W/2$ the other two states with ${\rm Im}\,k_{ss'}<0$ appear. In total $\psi$ has eight unknown coefficients, which we determine by demanding continuity of $\psi$ and $d\psi/dy$ at $y=W/2$ and $y=-W/2$. The determinant of this set of equations should vanish, in order to have a nontrivial solution. There is only a zero-mode for $U_{0}>0$, $U_{0}+\delta U<-m\Delta^{2}/2$, determined by
\begin{equation}
\tan qW=\frac{2qq_{0}}{q^{2}-q_{0}^{2}}.\label{qWrelation}
\end{equation}
We have defined
\begin{align}
&q=(m/\hbar)\sqrt{-\Delta^{2}-(2/m)(U_{0}+\delta U)},\label{qdef}\\
&q_{0}=(m/\hbar)\sqrt{\Delta^{2}+2U_{0}/m}.\label{q0def}
\end{align}
The MS states exist in between subsequent gap closings, as indicated in Fig.\ \ref{fig_cont} (shaded regions). 

\end{document}